# Singular evanescent wave resonances in moving media

Yu Guo and Zubin Jacob*

Department of Electrical and Computer Engineering,

University of Alberta, Edmonton, Alberta, T6G 2V4, Canada

*zjacob@ualberta.ca

**Abstract**

Resonators fold the path of light by reflections leading to a phase balance and thus constructive addition of propagating waves. However, amplitude decrease of these waves due to incomplete reflection or material absorption leads to a finite quality factor of all resonances. Here we report on our discovery that evanescent waves can lead to a perfect phase and amplitude balance causing an ideal Fabry-Perot resonance condition in spite of material absorption and non-ideal reflectivities. This counterintuitive resonance occurs if and only if the metallic Fabry-Perot plates are in relative motion to each other separated by a critical distance. We show that the energy needed to approach the resonance arises from the conversion of the mechanical energy of motion to electromagnetic energy. The phenomenon is similar to lasing where the losses in the cavity resonance are exactly compensated by optical gain media instead of mechanical motion. Nonlinearities and non-localities in material response will inevitably curtail any singularities however we show the giant enhancement in non-equilibrium phenomena due to such resonances in moving media.

**Conventional Fabry-Perot resonance**

The canonical example of a resonator is the Fabry-Perot (FP) system consisting of two reflecting plates separated by a vacuum gap [1,2]. Light bouncing between them serves as a textbook introduction to the concept of a resonance and is the basis of practical devices from the laser to the interferometer [1,2]. A simple argument suffices to understand this resonance. The reflection coefficient of propagating waves with frequency $\omega$ from the first mirror ($r_1(\omega)$) times that of the second mirror ($r_2(\omega)$) along with the propagation phase accumulated over a round trip ($e^{2ik_z d}$) should reconstruct the wave, capturing it inside, leading to a resonant build-up of intensity. Here, d is the vacuum gap between the mirrors and $k_z$ is the propagation constant perpendicular to the mirrors. We arrive at the Fabry-Perot resonance condition

$$r_1(\omega) r_2(\omega) e^{2ik_z d} = 1, \tag{1}$$

which also follows from a plane wave multiple scattering approach.

It is well known that this above equation cannot be fulfilled by any passive media. Note that the reflection coefficients are complex signifying the change in phase and amplitude of the propagating wave at the mirrors. A closer look reveals that an optimum choice of the gap can possibly lead to a net phase balance ($\arg(r_1(\omega) r_2(\omega) e^{2ik_z d}) = 2n\pi$) for a resonance, but material absorption and non-ideal reflections necessarily require $|r_1(\omega) r_2(\omega)| < 1$. A gain medium is needed to compensate for this loss in amplitude as in a laser. The arguments presented above can be generalized to arbitrary passive structures showing that the bound resonances are signified by the poles of the scattering matrix which always lie in the lower half ($\text{Im}(\omega_{res}) < 0$) of the complex frequency plane [3]. This condition ensures that all resonances decay in time leading to a finite quality factor.

In this paper, we show that the conventional Fabry-Perot condition has fundamental differences in the case of moving media. We explain that evanescent waves bouncing between moving plates can lead to a counterintuitive resonance with perfect amplitude and phase balance. We introduce the concept of a negative Poynting vector flow arising from Doppler shifted negative frequency modes in moving media. The spontaneous emission of negative frequency modes from the moving plate forms the subtle reason for the existence of such a resonance, which in essence is similar to the concept of a laser. The gain is provided by the conversion of the mechanical energy of motion into electromagnetic energy through the negative frequency modes. Finally we show that this resonance can dominate the non-equilibrium heat transfer leading to a singularity for a critical velocity of the moving plate. The singularities will inevitably be curtailed by nonlinearities and non-localities close to the resonance and we discuss in detail the effect of hydrodynamic non-locality on our predicted resonance.

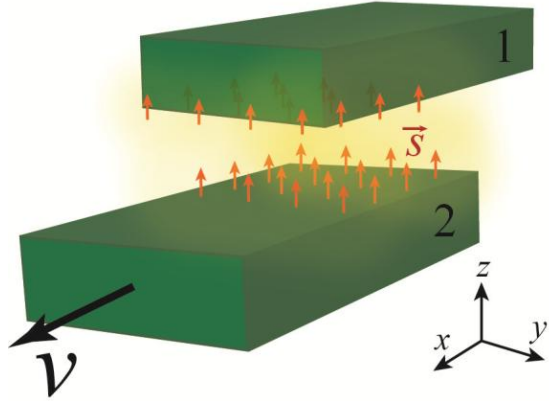

Figure 1. Singular Fabry-Perot (FP) resonance of evanescent waves can be achieved by setting the FP plates in relative motion. Plate 1 is stationary while plate 2 is moving at a constant velocity V along the x direction. The reflection coefficients and the distance for the moving case can lead to a perfect balance of both phase and amplitude which cannot occur for stationary plates.

**Singular resonance in moving Fabry-Perot plates**

We first provide an intuitive understanding of our phenomenon through arguments of phase and amplitude of reflection from moving media.

**i) Perfect phase balance for evanescent waves**

We consider evanescent waves bouncing between two plates. To achieve net round trip phase advancement of zero, we need the phase of reflection from the two plates to cancel each other. Note evanescent waves show no phase advancement within the gap, i.e., $e^{2ik_z d} = e^{-2|k_z|d}$ is purely real for evanescent waves. The simple condition which can achieve this is given by

$$r_2(\omega) = r_1^*(\omega) \tag{2}$$

This fulfills the phase balance condition for evanescent waves at the mirrors. Note a perfect phase balance of evanescent waves is fundamentally impossible in the conventional Fabry-Perot system. The full proof will be elaborated elsewhere.

Note that for any medium we have the condition $r^*(\omega) = r(-\omega)$ due to the condition on the reality of the fields [3]. This implies the complex conjugation of the reflection coefficient ($r^*(\omega)$) for evanescent wave Fabry-Perot resonances can be obtained by considering its negative frequency counterpart ($r(-\omega)$). Thus the problem of achieving such a resonance reduces to transforming the reflection from the second mirror into the negative frequency reflection coefficient of the first mirror

$$r_2(\omega) = r_1(-\omega) \qquad (3)$$

This can be achieved by setting two plates made of the same material in relative motion but with a fixed gap between them (Fig. 1)). For observers or bodies in relative motion, no concept of a single frequency exists. This is because the phase of the wave (a scalar) is the invariant of motion not the frequency (energy) or wavevector (momentum). Therefore it is possible for the two plates to perceive frequencies of opposite sign. Consider an incident plane wave in vacuum given by $e^{i(\vec{k}\cdot\vec{r}-\omega t)}$ on a moving interface. We consider the lateral wavevector $(k_x, k_y)$ along the non-relativistic moving direction ($\beta = V/c \ll 1$ and $k_y = 0$). The frequency of the wave in the frame co-moving at a constant velocity V along the x axis is Doppler shifted [2] to $\omega' = \omega - k_x V$. Note the reflection from a transversely moving plate does not alter the incident frequencies and can be expressed as the standard Fresnel reflection from a stationary plate with a Doppler shifted frequency, that is, $r^{mov}(\omega) = r(\omega')$.

This shifted frequency appears negative ($\omega' = \omega - k_x V < 0$) to the second plate for waves with $k_x > \omega/V$. Such waves have large wavevectors lying beyond the light line ($k_x \gg \omega/c$) and are necessarily evanescent in the gap. We conclude that an evanescent wave with frequency $\omega$ incident on the stationary plate 1 will appear Doppler shifted to $-\omega$ for the moving plate 2 when $-\omega = \omega - k_x V$. We call this the phase balance wavevector,

$$k_x^{PB} = 2\frac{\omega}{V} = \frac{2k_0}{\beta} \qquad (\omega' = -\omega) \tag{4}$$

where $k_0 = \omega/c$ is the free space wavevector and $\beta = V/c$. Evanescent waves with this special wavevector will bounce off the stationary first mirror with reflection coefficient $r_1(\omega)$ but reflect off the second identical but moving mirror with coefficient $r_2^{mov}(\omega) = r_1(-\omega) = r_1^*(\omega)$. Then, at the phase balance wavevector, the resonance condition Eq. 1 evolves into a new form,

$$r_1(\omega)r_2^{mov}(\omega)e^{2ik_z d} = r_1(\omega)r_1^*(\omega)e^{-2|k_z|d} = |r_1(\omega)|^2 e^{-2|k_z|d} = 1 \;. \tag{5}$$

We emphasize that perfect phase balance occurs since

$$\arg(r_1(\omega)) + \arg(r_1^*(\omega)) + \arg(e^{-2|k_z|d}) = 0, \tag{6}$$

which is valid inspite of losses and dispersion in the reflection coefficient.

## ii) Perfect amplitude balance for evanescent waves

In addition to perfect phase balance, we need to achieve perfect amplitude balance by compensating for the evanescent wave exponential decay in the gap between the moving plates. Those evanescent waves which couple to the surface modes of the mirror can have a reflection coefficient with amplitude greater than unity. Such waves can thus have $|r_1(\omega)r_2(\omega)| > 1$ to compensate the evanescent decay within the gap as well as non-ideal mirror reflections. Note the phase balance wavevector ($k_x^{PB} = 2k_0/\beta$) is much larger than $k_0$ so the reflection coefficient for p-polarized waves can be approximated by $r_p(\omega) = (\varepsilon(\omega) - 1)/(\varepsilon(\omega) + 1)$. When $\mathrm{Re}(\varepsilon(\omega_{SPR})) = -1$, there occurs a pole of the reflection coefficient corresponding to the surface plasmon resonance (SPR) [4] and we have

$$|r_p(\omega_{SPR})| > 1. \tag{7}$$

## iii) Singular condition

Now we discuss how to achieve this phase and amplitude balance (PAB) together to achieve an ideal Fabry-Perot condition. If you consider moving plates that support surface plasmon resonances, a special condition arises when the moving plate perceives a Doppler shifted negative SPR frequency $\omega' = -\omega_{SPR}$. This occurs at the SPR phase balance wavevector

$$k_x^{PB} = 2\frac{\omega_{SPR}}{V} \tag{8}$$

This leads to the Fabry-Perot resonance condition for evanescent waves

$$|r_1(\omega_{SPR})|^2 e^{-4\omega_{SPR}d/V} = 1 \tag{9}$$

where we have used $k_z = ik_x^{PB} = i2\omega/V$ as $k_x^{PB}$ is much larger than free space wavevector. Simultaneously amplitude balance arises since the exponential decay ($e^{-2|k_z|d}$) is compensated by the enhancement ($|r_p(\omega)|>1$) due to evanescent coupling with surface waves. The amplitude enhancement of an evanescent wave has a maximum at the SPR frequency $\omega_{SPR}$, leading to the critical distance $d_0 = V/(2\omega_{SPR})\ln|r_p(\omega_{SPR})|$ for a fixed velocity. Alternatively we can also interpret the above equation as the existence of a critical velocity

$$V_0 = \frac{2\omega_{SPR}}{d \ln|r_p(\omega_{SPR})|} \tag{10}$$

for a fixed gap ($d \ll \lambda$) which achieves the perfect amplitude balance. We adopt this viewpoint in all our calculations to emphasize that motion is the essential ingredient which causes the singularity.

### iv) Fundamental difference from stationary plates

We now emphasize the fundamental significance of phase balance by showing the difference between the stationary plate [5] and moving plate cases.

For evanescent waves, two identical stationary plates would lead to the well-known condition found in textbooks,

$$r^2 e^{-2|k_z|d} = 1 \quad \text{(impossible in stationary plates)} \tag{11}$$

As explained previously, the complex nature of the reflection coefficient implies this condition cannot be fulfilled irrespective of distance or material properties. The subtle balance of phase is completely missed by the above equation.

At the phase balance wavevector, the moving plate case gives rise to a counterintuitive singular resonance condition

$$|r|^2 e^{-2|k_z|d} = 1 \quad \text{(possible in moving plates)} \tag{12}$$

Note that only the amplitude of the reflection coefficient enters the moving Fabry-Perot condition. The subtle role of the negative frequency mode is revealed in the phase cancellation on reflection which cannot occur for stationary plates. We emphasize that this condition holds true in the relativistic case as well leading to a singular resonant condition in spite of the presence of material dispersion and absorption. A detailed proof of this result taking into account polarization mixing will be given elsewhere.

**Negative Frequency Photonic Modes**

We now introduce the concept of negative Poynting vector flow which only occurs for negative frequency modes. This negative behavior of conventionally positive quantities is completely intuitive since it is a manifestation of gain: conversion of mechanical energy of motion to electromagnetic energy through negative frequency photonic modes.

The energy of incident evanescent waves on any medium is given by the normal component of the Poynting vector

$$S_z = \frac{1}{2}\text{Re}(E \times H^*)_z \propto \frac{|k_z|}{\omega} 2\,\text{Im}(r). \tag{13}$$

For passive and stationary media, the energy tunneling into the medium is always positive ( $S_z > 0$, $\text{Im}(r) > 0$ ). There is a stark contrast between the Poynting vector directions for positive frequencies and negative frequencies (evanescent waves with $k_x > \omega/V$ ). Since $r(-\omega) = r^*(\omega)$, we see that the Poynting vector of tunneling negative frequency waves ( $S_z \propto \text{Im}(r) < 0$) is opposite to that of positive frequency waves. We thus note that as opposed to the conventional case of light tunneling into a medium, the negative frequency evanescent modes supported by a moving medium can tunnel out of it (Fig. 1). This is a manifestation of gain. These photons tunneling out of the moving plate will subsequently be absorbed by the stationary plate. Thus the existence of a singular resonance should come as no surprise since mechanical energy is being converted to electromagnetic energy.

**Radiative Heat transfer**

We now consider the effect of such a resonance on physical observables. We choose to study the radiative heat transfer between moving plates since this is a near-field phenomenon which has been experimentally studied with stationary metallic plates. We emphasize that the role of this resonance can play an important role in various phenomena.

**i) Material parameters, velocity and gap size**

To analyze the nature of the excited evanescent wave resonance in a practical scenario, we consider two identical metallic plates moving relative to each other at non-relativistic speeds. The moving velocity ($V$) is bounded by the phonon velocity of the medium [6], typically in the order of $10^4$m/s, thus $\beta = V/c \ll 1$. The plates are separated by a small gap to allow for interaction through large wavevector evanescent waves, necessary to achieve Doppler shifted negative frequencies in the co-moving frame. At the surface wave resonance frequency, the magnitude of reflection coefficient $|r_p|$ is bounded by the loss of the material. Realistic estimates for $|r_p|$ is in the order of 10 and if the gap size is 10nm, the operating frequency $\omega_{SWR}$ should be in the order of $10^{12}$Hz (1THz) to give a critical velocity in the order of $10^4$m/s. Thus we need materials with low plasma frequency in the THz region or even lower to observe the singular Fabry-Perot resonance of evanescent waves. THz surface waves have been generated in in degenerately doped semiconductors [7], phonon-polaritonic polar dielectrics, low frequency plasmons [11] and graphene [9]. Here we consider a Drude metal with frequency dependent permittivity given by $\varepsilon(\omega) = 1 - 2\omega_{SPR}^2/(\omega^2 + i\Gamma\omega)$, where the surface plasmon resonance frequency $\omega_{SPR} = 10^{12}$Hz and the loss factor $\Gamma = 0.01\omega_{SPR}$. In Fig. 2, we plot the reflection coefficients from both stationary and moving plates made of this metal and demonstrate the concept of PAB condition.

At the gap width of $d = 20$nm, the critical velocity that satisfies the singular resonance condition is $V_0 \approx 8.7 \times 10^3$, and the phase balance wavevector $k_x^{PB}$ is in the order of $10^4 k_0$. Note that we do take into account material dispersion and absorption and discuss the role of non-locality [10,11] due to large wavevectors in the final section.

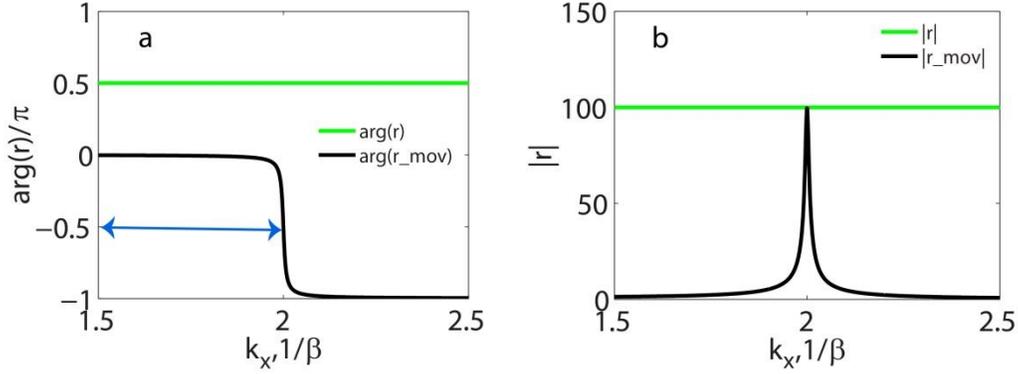

Figure 2. Phase (a) and amplitude (b) of reflection coefficients at the SPR frequency from stationary and moving plates. The velocity of the moving plate is $10^4$m/s. For a stationary plate (green curves), the reflection coefficient is almost a constant at these huge wavevectors. However, the frequencies perceived by the moving plates are negative, ranging from $\omega'=-0.5\omega_{SPR}$ to $\omega'=-1.5\omega_{SPR}$, so the phases of the reflection coefficients are negative. Furthermore, the phase at the phase balance wavevector is exactly opposite to that of the stationary plate (see blue arrow). In (b), we can clearly see that the amplitude of r can be larger than unity which is essential for amplitude balance. The amplitude for the moving plate has a unique dispersion which peaks at the phase balance wavevector where $\omega'=-\omega_{SPR}$.

**ii) Fundamental difference in the multi-reflection factor**

Rytov's fluctuational electrodynamics gives the number of evanescent photons exchanged between the plates with frequency $\omega$ and lateral wavevector $(k_x, k_y)$ as [12–15]

$$N(\omega, k_x, k_y) = \frac{2\operatorname{Im}[r_1(\omega)]|e^{ik_z d}|^2 \, 2\operatorname{Im}[r_2^{mov}(\omega)](n(\omega', T_2) - n(\omega, T_1))}{|1 - r_1(\omega)r_2^{mov}(\omega)e^{2ik_z d}|^2} \quad . \tag{14}$$

Here $n(\omega, T)$ is the Bose-Einstein occupation number. In the calculations we choose the temperatures of the plates to be $T_1$=320K and $T_2$=300K. Note that the occupation number of the moving plate should be evaluated at the Doppler shifted frequency [12]. $r_1(\omega)$ and $r_2^{mov}(\omega)$ are the reflection coefficients for the stationary and moving plates respectively evaluated for a wave incident with frequency $\omega$ in the lab frame. We have $r_2^{mov}(\omega) = r_1(\omega')$ and the factor $|e^{ik_z d}|^2$ accounts for the decay of the photon propagating between the two plates while $1/|1-r_1(\omega)r_2^{mov}(\omega)e^{2ik_z d}|^2$ is for the multi-reflection between the plates.

Even though such multi-reflection factors are routinely encountered in the case of parallel plates, once they are set in motion we predict a fundamental difference. For frequencies at which the plates support surface waves, the singular Fabry-Perot resonance of evanescent waves can lead to the divergence of this multi-reflection factor for the phase balance wavevector ($k_x^{PB} = 2\omega_{SPR}/V$) and critical velocity. The role of evanescent waves and surface waves in the mediation of energy transfer as well as Casimir forces are well known [5,16]. However, it should be noted that these conventional resonant modes which are essentially poles of the scattering matrix always lie in the lower half of the complex plane for passive media with finite imaginary part of resonant frequency. Our central claim is the remarkable fact that motion of the plate pulls the pole up to the real axis leading to a real solution for the resonant frequency.

### iii) Photon transfer

In Fig. 3, we plot the spectrum of photons exchanged according to their frequency and wavevector in the lab frame. For a distance $V_1$ which is away from the singular Fabry-Perot Resonance condition, we see two distinct bright lines in $\omega - k$ space through which photons are exchanged between the two plates. The horizontal line corresponds to the SPR frequency of the stationary plate where all wavevectors are excited like in a conventional surface plasmon resonance [17]. Note only the large wavevectors far from the light line have been shown. The oblique line corresponds to the Doppler shifted SPR frequency of the moving plate.

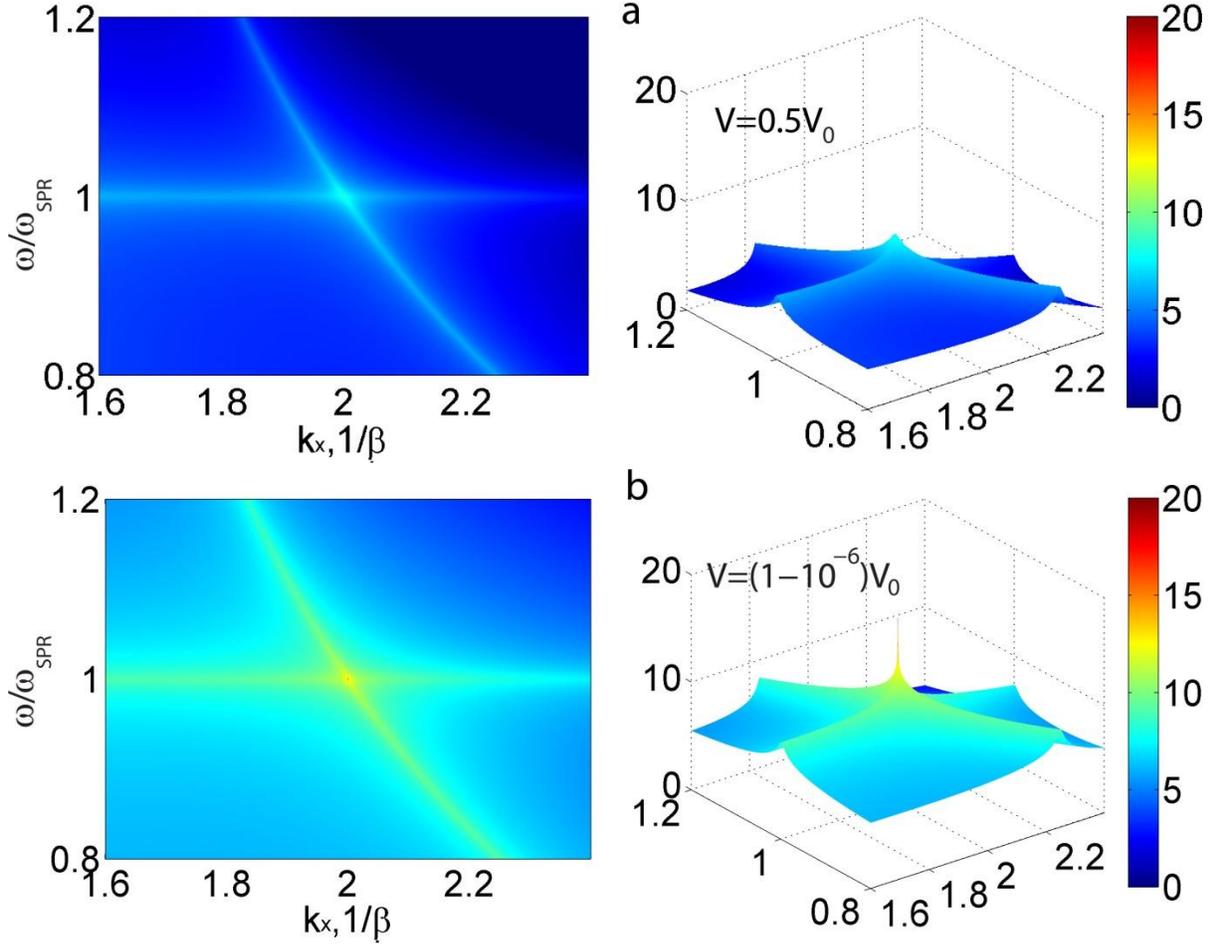

Figure 3. Contribution to exchanged photon number resolved by frequency and lateral wavevector $k_x$ (normalized to free space wavevector) at (a) $V=0.5V_0$ and (b) $V \to V_0$. The right panels are 3-dimensional plots. In both (a) and (b), we see two bright curves, both of which are due to surface plasmon resonances. In (a), at a velocity away from $V_0$, the singular resonance condition is far from being satisfied. However, the bright curves remain due to the SPR at the two interfaces. In (b), the large peak in the middle is due to the singular resonance that arises since the amplitude balance condition is satisfied when $V \to V_0$ and phase balance condition is satisfied at $k_x=2/\beta$. This leads to giant photon exchange between moving plates at phase and amplitude balance (PAB) condition.

Our result shows that as the velocity of the moving plate is increased closer to the singular FP resonance condition ($V \to V_0$), a fundamentally new mechanism of photon exchange emerges. This is evident from Fig. 3b where photons with the phase balance wavevector completely dominate the interaction near the amplitude balance velocity. Note that this occurs when the frequencies in the co-moving frame and lab frame are exactly opposite, the condition for phase balance. Indeed, the multiple scattering term $1 - r_1 r_2 e^{2ik_z d}$ in Eq. 14, vanishes giving rise to an infinitely large number of photons exchanged when PAB condition is fulfilled.

**iv) Giant heat transfer and Critical velocity**

We assert that the singular evanescent wave resonance fundamentally dominates all non-equilibrium processes between the plates. We focus here on the radiative heat transfer between the moving plates, which is the product of the total number of photons exchanged (Eq. 14) and the energy of a single photon $\hbar\omega$ giving $h(\omega, k_x, k_y) = \hbar\omega N(\omega, k_x, k_y)$. The net heat transfer is expressed by integrating all the partial waves,

$$H = \int \frac{d\omega}{2\pi} \int \frac{dk_x}{2\pi} \frac{dk_y}{2\pi} h(\omega, k_x, k_y) \tag{15}$$

In Fig. 4a, we plot the spectrum of this transferred energy resolved according to the wavevector for plate velocities close to the critical velocity $V = V_0^-$ and $V = 0.5V_0$. The largest contribution to the heat transfer is due to the singular Fabry-Perot resonance of evanescent waves. This is discerned from the fact as the velocity of the moving plate increases, a dominant contribution occurs at the phase balance wavevector. Close to the critical velocity, we predict the non-equilibrium heat transfer to be $H \sim \ln(V_0 / (V_0 - V))$. We plot the friction vs. distance in Fig. 4b to verify the theoretical scaling law. We clearly see that the friction increases as $\ln[V_0/(V_0 - V)]$ when $V$ approaches $V_0$.

We give an estimate of the giant enhancement in the heat transfer as the singular resonance is approached. In Fig. 4b, the magnitudes of heat transfer evaluated around the resonance at $V_1 = 0.5V_0$ and $V_2 = (1-10^{-6})V_0$ are 0.066W/m$^2$ and $7.2 \times 10^3$W/m$^2$, respectively. We emphasize the large order of magnitude increase in the heat transfer as the critical velocity is reached showing the important role of the resonance.

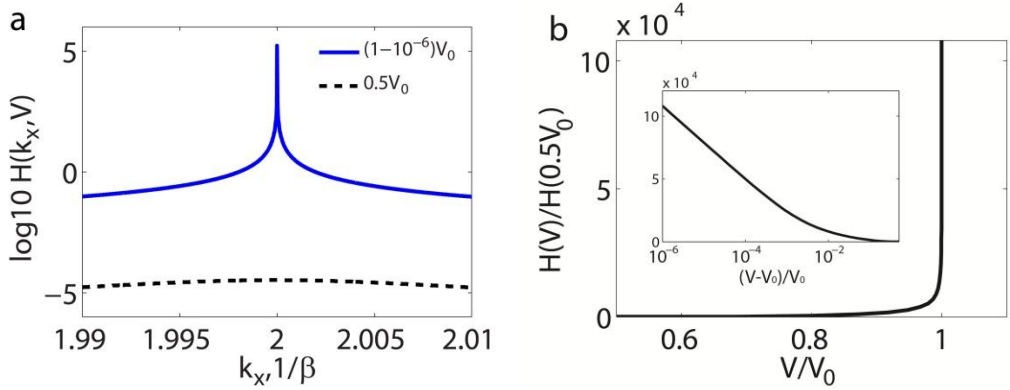

Figure 4. Heat transfer on the moving plates (a) resolved by the wavevector $k_x$ at $V \rightarrow V_0$ and $V=0.5V_0$. A major contribution to the heat arises from modes at the perfect phase balance wavevector. Note, at $0.5V_0$, the amplitude of friction is significantly smaller. (b) The distance velocity of heat transfer at velocities near $V_0$. It is very clear that heat transfer goes up rapidly near the critical velocity $V_0$. In the inset, the x axis is $(V_0-V)/V_0$ and in log scale. We clearly see a linear increasing behavior as V approaches $V_0$. This is consistent with the theoretical scaling law which predicts a giant heat transfer.

## v) Role of non-locality

We now discuss effects which will inevitable curtail any singularities in physical observables. We do not assume ideal mirrors [18] and losses or dispersion are not a fundamental impediment to the singular resonance. The phase balance condition also occurs irrespective of losses or dispersion. However, the amplitude balance condition relies on the enhancement of evanescent waves at the surface plasmon resonance. This can be damped if the losses are too large. Another important factor is that the non-local response of any metal at large wavevectors can also reduce the effective enhancement. To analyze this effect, we consider a hydrodynamic non-locality [11] with permittivity $\varepsilon(\omega,k) = 1 - 2\omega_{SPR}^2/(\omega^2 + i\Gamma\omega - \alpha^2 k^2 c^2)$, where α is the non-locality parameter. Using additional boundary conditions, we evaluate the reflection coefficient for a spatially dispersive metal. Fig. 5 shows the amplitude enhancement factor as a function of the non-locality parameter which depends on the Fermi velocity of the electrons in the metal. Note for velocities in the range of $\beta = 10^{-4}$, the phase balance wavevector is $2 \times 10^4 k_0$. Surface waves with such large wavevectors are damped as the non-locality parameter increases. The non-locality parameter α practical metal is in the order of $10^{-3}$. In Fig. 5a, we clearly see that at this velocity $|r|$ is smaller than 1 for practical non-locality parameter, which means that the amplitude balance cannot be fulfilled. However, it is important to note the remarkable fact that if media start moving at velocities comparable to the Fermi velocity of electrons in the order of $\beta = 10^{-2}$, we recover the local theory since the phase balance wavevector is now $200k_0$, where the non-local effect is much smaller than that at the former wavevector. At this larger velocity, the amplitude of reflection can be larger than 1 taking non-locality into account (see Fig. 5b). Material nonlinearities will be another important effect which will curtail any singularities near this resonance. A detailed analysis of material non-linearities and non-locality will be provided elsewhere.

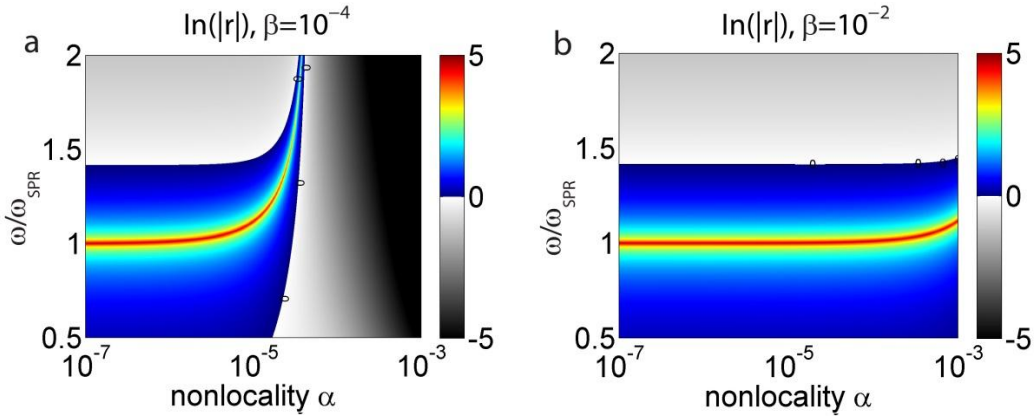

Figure 5. Amplitude of reflection coefficients at various velocities and non-locality parameters. The red and blue regions denote that amplitudes are larger than unity, where amplitude balance is possible, while gray region denotes amplitudes less than unity which cannot satisfy amplitude balance. We emphasize that at a higher velocity comparable to the Fermi velocity of the electrons, the amplitude can be larger than one with non-locality taken into account.

**Conclusion**

In conclusion, we have reported about a singular Fabry-Perot resonance which can never occur in the conventional case of stationary plates. This can be understood as a perfect coupling of positive and negative frequencies in the near-field and can find widespread applicability in the enhancement of many negative frequency related effects (eg: quantum friction). One of the main challenges in observing the phenomena is the physical motion of plates at high velocities. However, we believe light induced potentials and other innovative optomechanical approaches will be the true testing ground of our predicted phenomenon. Furthermore, the subtle concept of phase and amplitude balance we have introduced can lead to a fundamental understanding of negative frequency photonic modes.

**Acknowledgments:**
We acknowledge funding from National Science and Engineering Research Council of Canada, Helmholtz Alberta Initiative and Alberta Innovates Technology Futures.

**References:**
1. M. Wolf and E. Born, *Principles of Optics: Electromagnetic Theory of Propagation, Interference and Diffraction of Light* (Cambridge University Press, 1980).
2. J. A. Kong, *Electromagnetic Wave Theory* (Wiley New York, 1990), Vol. 2.
3. L. Landau, E. Lifshitz, and L. Pitaevskii, *Electrodynamics of Continuous Media* (Pergamon Press, Oxford, 1984).
4. H. Raether, *Surface Plasmons on Smooth Surfaces* (Springer, 1988).
5. F. Intravaia and A. Lambrecht, "Surface Plasmon Modes and the Casimir Energy," Phys. Rev. Lett. **94**, 110404 (2005).
6. J. B. Pendry, "Shearing the vacuum - quantum friction," J. Phys. Condens. Matter **9**, 10301 (1997).
7. A. J. Hoffman, L. Alekseyev, S. S. Howard, K. J. Franz, D. Wasserman, V. A. Podolskiy, E. E. Narimanov, D. L. Sivco, and C. Gmachl, "Negative refraction in semiconductor metamaterials," Nat. Mater. **6**, 946–950 (2007).
8. R. Zhao, A. Manjavacas, F. J. García de Abajo, and J. B. Pendry, "Rotational Quantum Friction," Phys. Rev. Lett. **109**, 123604 (2012).
9. A. I. Volokitin and B. N. J. Persson, "Quantum Friction," Phys. Rev. Lett. **106**, 094502 (2011).
10. C. Ciracì, R. T. Hill, J. J. Mock, Y. Urzhumov, A. I. Fernández-Domínguez, S. A. Maier, J. B. Pendry, A. Chilkoti, and D. R. Smith, "Probing the Ultimate Limits of Plasmonic Enhancement," Science **337**, 1072–1074 (2012).
11. A. Moreau, C. Ciracì, and D. R. Smith, "Impact of nonlocal response on metallodielectric multilayers and optical patch antennas," Phys. Rev. B **87**, 045401 (2013).
12. M. F. Maghrebi, R. Golestanian, and M. Kardar, "Scattering approach to the dynamical Casimir effect," Phys. Rev. D **87**, 025016 (2013).
13. S.-A. Biehs, E. Rousseau, and J.-J. Greffet, "Mesoscopic Description of Radiative Heat Transfer at the Nanoscale," Phys. Rev. Lett. **105**, 234301 (2010).
14. A. I. Volokitin and B. N. J. Persson, "Theory of the interaction forces and the radiative heat transfer between moving bodies," Phys. Rev. B **78**, 155437 (2008).
15. J. B. Pendry, "Radiative exchange of heat between nanostructures," J. Phys. Condens. Matter **11**, 6621 (1999).
16. S. Shen, A. Narayanaswamy, and G. Chen, "Surface Phonon Polaritons Mediated Energy Transfer between Nanoscale Gaps," Nano Lett. **9**, 2909–2913 (2009).
17. L. Novotny and B. Hecht, *Principles of Nano-Optics* (Cambridge university press, 2006).
18. M. Kardar and R. Golestanian, "The "friction" of vacuum, and other fluctuation-induced forces," Rev. Mod. Phys. **71**, 1233–1245 (1999).